\documentclass[%
 reprint,
 amsmath,amssymb,
 aps,
]{revtex4-2}

\usepackage{graphicx}
\usepackage{dcolumn}
\usepackage{bm}
\usepackage{tikz}
\usetikzlibrary{positioning, shapes.geometric}
\usepackage{epstopdf}
\usepackage{hyperref}

\graphicspath{{./}{Figures/}}

\begin{document}

\preprint{APS/123-QED}

\title{Periodic Cylindrical Bilayers Self-Assembled from Diblock Polymers}

\author{Yongshun Luo}
\affiliation{School of Mathematical Sciences, Beijing Normal University, Beijing 100875, China.}
\author{Min Yang}
\affiliation{School of Mathematics and Statistics, Guizhou University, Guiyang 550025, China.}

\author{Sirui Li}
\affiliation{School of Mathematics and Statistics, Guizhou University, Guiyang 550025, China.}%
\author{Yana Di}
\email{yndi@uic.edu.cn}
\affiliation{Research Center for Mathematics, Beijing Normal University, Zhuhai 519087, China.}%
\affiliation{Guangdong Key Laboratory of IRADS, BNU-HKBU United International College, Zhuhai 519087, China.}
\author{Yongqiang Cai}
\email{caiyq.math@bnu.edu.cn}
\affiliation{School of Mathematical Sciences, Beijing Normal University, Beijing 100875, China. }
\date{\today}

\begin{abstract}
Amphiphilic polymers in aqueous solutions can self-assemble to form bilayer membranes, and their elastic properties can be captured by the well-known Helfrich model involving several elastic constants. In this paper, we employ the self-consistent field model to simulate sinusoidal bilayers self-assembled from diblock copolymers where a proper constraint term is introduced to stabilize periodic bilayers with prescribed amplitudes. Then, we devise several methods to extract the shape of these bilayers and examine the accuracy of the free energy predicted by the Helfrich model. Numerical results show that when the bilayer curvature is small, the Helfrich model predicts the excess free energy more accurately. However, when the curvature is large, the accuracy heavily depends on the method used to determine the shape of the bilayer. In addition, the dependence of free energy on interaction strength, constraint amplitude, and constraint period are systematically studied. Moreover, we obtain certain periodic cylindrical bilayers that are equilibrium states of the self-consistent field model, which agree with the theoretical predictions made by the shape equations.
\begin{description}
\item[Keywords]
Self-consistent field theory; Helfrich model; flexible membrane.
\end{description}
\end{abstract}
\maketitle

\section{Introduction}

Bilayer membranes are widely found in biological and chemical systems, including cell membranes, nuclei, chloroplasts, and mitochondria \cite{alberts2017molecular, 10.1104/pp.109.149070, korpelainen2004evolutionary}. Its unique bilayer structure, composed of hydrophilic and hydrophobic molecules, makes it have broad applications in biomedicine and biotechnology, such as drug delivery carriers, nanomaterial preparation, biological imaging, energy storage and conversion \cite{Hamley2004Self, Bates2012Multiblock, 2016Bioinspired}. To gain a deeper understanding and utilization of their structure and properties, polymeric hybrid membranes are broadly studied to mimic biological membranes because there are virtually no limits to the selection of monomers and chain architecture \cite{Taubert2004Self}.

In recent decades, the increasing computing power has enabled individuals to analyze and simulate research objects at a deeper, more detailed, and more comprehensive level. In particular, there are three frameworks for membrane simulations at different scales: (1) particle level methods, including molecular dynamics, Monte Carlo, and dissipative particle dynamics \cite{1995In, pivkin2010dissipative, 1998Dissipative}; (2) field theory methods, such as self-consistent field (SCF) theory \cite{2007Selfcc, Fredrickson2006Equilibrium} and density functional theory \cite{uneyama2005density}; (3) surface-based methods, such as Helfrich curvature elastic theory and Hamm-Kozlov model \cite{Helfrich1973Elastic, hamm1998tilt}. Among the various theoretical frameworks developed for amphiphilic molecules, the SCF theory provides a versatile and accurate framework for studying self-assembled bilayer membranes \cite{Li2013Elastic, Zhang2015Application}. However, if we neglect the inner structure and simplify the topological structure of the membrane as a surface $S$, the free energy of the membrane can be approximated by the Helfrich model which expressions the energy as \cite{Helfrich1973Elastic, Li2013Elastic}
\begin{align}
\label{eq:energy functional}
F=\int_{S}\{&\gamma+2 \kappa_{M}\left(M-c_{0}\right)^{2}
+\kappa_{G} G\nonumber\\
&+\kappa_{1} M^{4}+\kappa_{2} M^{2} G+\kappa_{3} G^{2}\}\mathrm{d}A.
\end{align}
where $M=(c+c')/2$ and $G=cc'$ are the local mean curvature and Gaussian curvature, respectively, and $c$ and $c'$ are the two principal curvatures. The parameters $\gamma$, $c_0$, $\kappa_M$, and $\kappa_G$ in the model are called elastic constants, which are the surface tension, spontaneous curvature, bending modulus, and Gaussian modulus, respectively. In addition, $\kappa_1$, $\kappa_2$, and $\kappa_3$ are the fourth-order moduli, which are not neglectable to predict free energy of bilayer membranes with large curvatures \cite{Li2013Elastic}.

It is valuable to establish the relationship between different simulation methods and clarify their applicable scope and accuracy. For the relationship between the SCF and Helfrich model, Laradji et al. \cite{Laradji1998Elastic} used the SCF model to study the elastic properties of monolayer membranes, Li et al. \cite{Li2013Elastic} extended the elastic properties to bilayer membranes which were further extended to liquid-crystalline bilayers \cite{Cai2019Elastic}. In these studies, the bilayer membranes are restricted to planar, cylindrical, and spherical geometries and aim to extract the elastic constants. Taking the curvature of cylindrical and spherical membranes as small parameters, Cai et al. \cite{Cai2020Elastic, Wxy2021Analytical} obtained the analytical expression for the elastic constants and found that there are some energy differences between the results of the SCF simulation and the predictions of the Helfrich model when the curvatures are large. For general membrane geometries, the disparity between the SCF and Helfrich model has been less studied.

In this paper, we investigate self-assembled periodic cylindrical bilayers and examine the accuracy of the Helfrich model to predict their free energy. Particularly, we focus on sinusoidal shape bilayers, which were observed in experiments by Harbich et al. \cite{harbich1984swelling} and explored theoretically by Ou-Yang et al. \cite{PhysRevE.53.4206, ou1999geometric}. Within the framework of the SCF theory, we design a suitable constraint term to stabilize the bilayers with the desired shape and then extract the interfaces of the obtained bilayers. Comparing the free energies of the SCF and Helfrich models, we have discovered that the accuracy of the Helfrich model heavily depends on the method used to determine the shape of the membranes. In addition, when the interface is extracted from parallel surfaces, the equilibrium shape of periodic cylindrical bilayers agrees with the theoretical prediction of Ou-Yang et al. \cite{PhysRevE.53.4206}.

The rest of the paper is organized as follows. Section~\ref{sec:Model} describes the SCF model of the bilayer and the geometric constraints. In Section~\ref{sec:Method}, we provide the numerical method for solving the SCF model and extracting the bilayer profile. The main results are shown in Section~\ref{sec:Results and discussion}. Finally, a summary is provided in Section~\ref{sec:Conclusion}.

\section{Model}
\label{sec:Model}

\subsection{Self-consistent field model}
\label{subsec:SCF model}

The molecular model considered in this paper is a binary mixture system composed of $AB$-diblock copolymers and $hA$-homopolymers, following \cite{Li2013Elastic}. In this general model, the $AB$-diblock copolymers are used to represent amphiphilic molecules, while the $hA$-homopolymers are used to represent amphiphilic solvent molecules. We assume that the $hA/AB$ mixture system is incompressible and that both monomers ($A$ and $B$) have the same monomeric density $\rho_{0}$. The volume fractions of $A$ and $B$-blocks in the copolymers are denoted by $f_A$ and $f_B=1-f_A$, respectively. For simplicity, we further assume that the $AB$ and $hA$ polymers have the same degrees of polymerization $N$, and $A, B$ monomers have the same statistical segment length. The strength of the mutual repulsion between the $A$ and $B$ monomers is characterized by the Flory-Huggins parameter $\chi N$ \cite{Flory1953Principles}. The last parameter, the chemical potential $\mu_c$ or the corresponding activity degree $z_{c}=\exp \left(\mu_{c}\right)$, is used to control the average concentration of $AB$ or $hA$ molecules. In the grand canonical ensemble \cite{Li2013Elastic, Laradji1998Elastic, Fredrickson2006Equilibrium}, the free energy of the $hA/AB$ system within the SCF framework is given by
\begin{equation}
\begin{aligned}
\frac{N \mathcal{F}}{k_{B} T \rho_{0}}&=\int\mathrm{d}\mathbf{r}\Big[\chi N \phi_{A}(\mathbf{r}) \phi_{B}(\mathbf{r})-\omega_{A}(\mathbf{r}) \phi_{A}(\mathbf{r})\\
&-\omega_{B}(\mathbf{r}) \phi_{B}(\mathbf{r})-\xi(\mathbf{r})(\phi_{A}(\mathbf{r})+\phi_{B}(\mathbf{r})-1)\\
&+I_{\psi_{i}}\Big] -z_{c} Q_{c}-Q_{h},\label{eq:free energy}
\end{aligned}
\end{equation}
where $\phi_\alpha(\mathbf{r})$ and $\omega_\alpha(\mathbf{r})$ are the local concentration and mean field of the $\alpha$-type monomers, respectively. The local pressure $\xi(\mathbf{r})$ is a Lagrange multiplier that maintains the incompressibility of the system, and $I_{\psi_{i}}$ is a proper term to stabilize bilayers with various geometries. In addition, $Q_c$ and $Q_h$ are the contributions from the single-chain partition functions of the $AB$ and $hA$ molecules, respectively.

Motivated by previous geometric constraints \cite{Matsen1994Stable, Matsen1999Elastic, Li2013Elastic, Cai2019Elastic} for cylindrical and spherical membranes, the term $I_{\psi_{i}}$ here is designed as the following,
\begin{align}
    \displaystyle I_{\psi_{i}}=\sum_{i=1}^{n_{CP}}\psi_{i}G_{\varepsilon}(\mathbf{r}-\mathbf{r}_{i})(\phi_{A}(\mathbf{r})-\phi_{B}(\mathbf{r})),
\end{align}
where $n_{CP}$ number of constraint points $\mathbf{r}_i$ introduced to guide the bilayer's shape, $\psi_{i}$ are Lagrange multipliers, and $G_{\varepsilon}\left(\mathbf{r}-\mathbf{r}_{i}\right)$ are sharp Gaussians used to ensure that $\psi_{i}$ only operates near the position $\mathbf{r}_{i}$.

In SCF theory, the fundamental quantities we need to calculate are the probability distribution functions of the polymers, \emph{i.e.}, the propagators $q^{h}_A(\mathbf{r},s)$ for $A$-homopolymers and $q^{\pm}_A(\mathbf{r},s),q^{\pm}_B(\mathbf{r},s)$ for $AB$-diblock copolymers. These propagators are obtained by solving the following modified diffusion equations (MDEs) for flexible polymer chains in mean fields $\omega_A$ and $\omega_B$,
\begin{align}
    &\frac{\partial}{\partial s} q_{A}^{h}(\mathbf{r}, s)
    =\left(\nabla_{\mathbf{r}}^{2}-\omega_{A}(\mathbf{r})\right) q_{A}^{h}(\mathbf{r}, s),
    \quad s \in\left(0, 1\right),
    \label{eq: propagator eqs1}\\
    &\frac{\partial}{\partial s} q_{\alpha}^{\pm}(\mathbf{r}, s)
    =\left(\nabla_{\mathbf{r}}^{2}-\omega_{\alpha}(\mathbf{r})\right) q_{\alpha}^{\pm}(\mathbf{r}, s),
    \quad s \in\left(0, f_{\alpha}\right),
    \label{eq: propagator eqs2}
\end{align}
with the initial conditions $q_{A}^{h}(\mathbf{r}, 0)=q_{A}^{-}(\mathbf{r}, 0)=q_{B}^{-}(\mathbf{r}, 0)=1,q_{A}^{+}(\mathbf{r}, 0)=q_{B}^{-}(\mathbf{r}, f_{B})$ and $q_{B}^{+}(\mathbf{r}, 0)=q_{A}^{-}(\mathbf{r}, f_{A})$. In terms of the chain propagators, the single-chain partition functions are given by $Q_{c}=\int\mathrm{d}\mathbf{r} q_{A}^{+}\left(\mathbf{r}, f_{A}\right)$,$Q_{h}=\int\mathrm{d}\mathbf{r} q_{A}^{h}\left(\mathbf{r}, 1\right)$. Furthermore, the local concentrations of the $A$ and $B$ monomers are obtained from the propagators as
\begin{align}
    \phi_{A}(\mathbf{r})=&\int_{0}^{1}\mathrm{d}s q_{A}^{h}(\mathbf{r}, s) q_{A}^{h}\left(\mathbf{r}, 1-s\right)\nonumber\\
    &+z_{c} \int_{0}^{f_{A}}\mathrm{d}s q_{A}^{-}(\mathbf{r}, s) q_{A}^{+}\left(\mathbf{r}, f_{A}-s\right),\\
    \phi_{B}(\mathbf{r})=&z_{c} \int_{0}^{f_{B}}\mathrm{d}s q_{B}^{-}(\mathbf{r}, s) q_{B}^{+}\left(\mathbf{r}, f_{B}-s\right).
\end{align}
The corresponding SCF equations become
\begin{gather}
    \label{eq: Discrete SCF equations 1}
    \omega_A(\mathbf{r})=\chi N \phi_B(\mathbf{r})-\xi(\mathbf{r})+\sum_{i=1}^{n_{CP}} \psi_i G_{\varepsilon}\left(\mathbf{r}-\mathbf{r}_i\right), \\
    \label{eq: Discrete SCF equations 2}
    \omega_B(\mathbf{r})=\chi N \phi_A(\mathbf{r})-\xi(\mathbf{r})-\sum_{i=1}^{n_{CP}} \psi_i G_{\varepsilon}\left(\mathbf{r}-\mathbf{r}_i\right), \\
    \label{eq: Discrete SCF equations 3}
    \phi_A(\mathbf{r})+\phi_B(\mathbf{r})=1, \\
    \label{eq: Discrete SCF equations 4}
    \int\mathrm{d}\mathbf{r} G_{\varepsilon}\left(\mathbf{r}-\mathbf{r}_i\right)\left(\phi_A(\mathbf{r})-\phi_B(\mathbf{r})\right)=0.
\end{gather}
Generally, the SCF equations are hard to be solved analytically and we turn to the numerical solutions by iterating methods. During the iteration, we need to assign the update rule of the Lagrange multiplier $\psi_{i}$. Combine Eq.~(\ref{eq: Discrete SCF equations 1}) and (\ref{eq: Discrete SCF equations 2}), we have
\begin{align}
    \label{eq:scf_transt1}
    \omega_A(\mathbf{r})-\omega_B(\mathbf{r})&+\chi N(\phi_A(\mathbf{r})-\phi_B(\mathbf{r}))\nonumber\\
    &
    =
    2 \sum_{i=1}^{n_{CP}} \psi_i G_{\varepsilon}\left(\mathbf{r}-\mathbf{r}_i\right),
\end{align}
Multiplied by $G_{\varepsilon}(\mathbf{r}-\mathbf{r}_j)$ in both side, integrated along $\mathbf{r}$, and combined with Eq.~(\ref{eq: Discrete SCF equations 4}), the Eq.~(\ref{eq:scf_transt1}) becomes
\begin{align}
    \sum_{i=1}^{n_{CP}}M_{ij}\psi_{i}
    :=
    &\sum_{i=1}^{n_{CP}}\psi_i\int_V G_{\varepsilon}(\mathbf{r}-\mathbf{r}_i)G_{\varepsilon}(\mathbf{r}-\mathbf{r}_j)\mathrm{d}\mathbf{r}\nonumber\\
    &
    =
    \frac{1}{2}\int_V (\omega_A(\mathbf{r})-\omega_B(\mathbf{r})) G_{\varepsilon}(\mathbf{r}-\mathbf{r}_j)\mathrm{d}\mathbf{r},\nonumber\\
    &
    =:
    b_{j},~\forall j.
\end{align}
This is a system of linear algebraic equations $M \psi=b$, which can be easily solved to obtain each $\psi_{i}$.

\subsection{Excess free energy and elastic model}
\label{subsec:Bulk phase, excess free energy and elastic model}

The SCF equations usually have multi-solutions except for the constant solutions, \emph{i.e.}, the bulk phases, which can be solved analytically \cite{Li2013Elastic,Laradji1998Elastic}. The bulk phase is the environment in which the bilayer exists, and its free energy $\mathcal{F}_{\text {bulk }}$ is taken as a reference to define the excess free energy of bilayer membranes. Since the free energy difference $\mathcal{F}-\mathcal{F}_{b u l k}$ is proportional to the area of the membrane $\mathcal{A}$, the excess free energy density can be defined as
\begin{equation}
    F_{e x}=\frac{N\left(\mathcal{F}-\mathcal{F}_{b u l k}\right)}{k_{B} T \rho_{0} \mathcal{A}},
\end{equation}
where the expressions of $\mathcal{F}_{\text {bulk }}$ can be referenced in Appendix~\ref{sec:Bulk phase}. Here, we emphasize that the excess free energy, $F_{ex}$, rather than the original energy $\mathcal{F}$ is compared with the energy predicted by the Helfrich model. To employ the Helfrich model, one should first determine the elastic constants. Previous studies have shown that simulating cylindrical and spherical bilayers with various curvatures is enough to extract the elastic constants of bilayers \cite{Li2013Elastic}, and this procedure can be improved by using the asymptotic expansion method to obtain analytical expressions of the elastic constants \cite{Cai2020Elastic}.

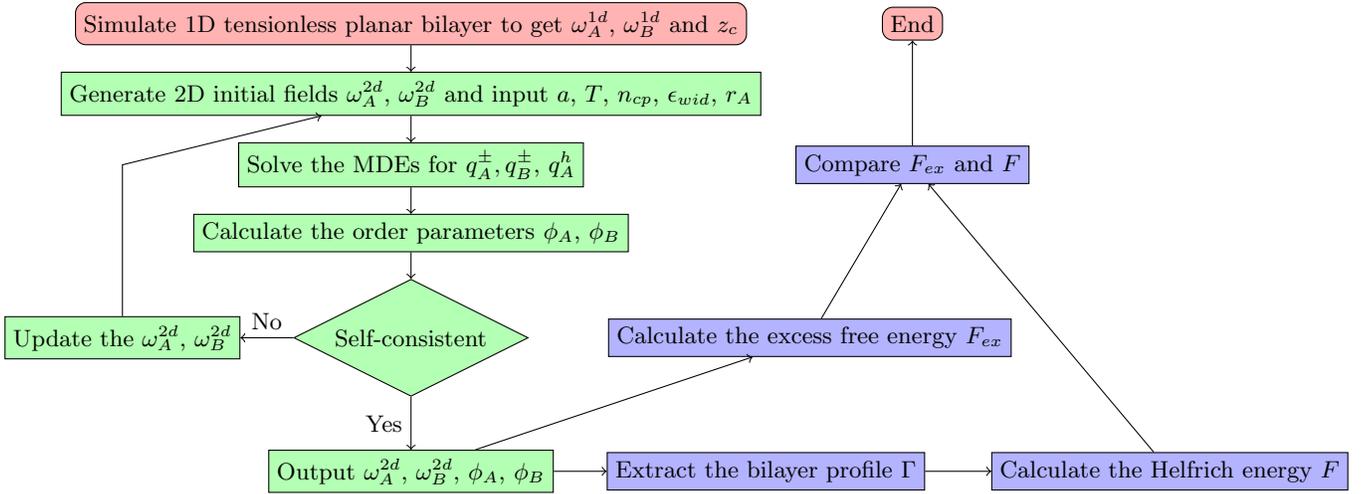
\begin{figure*}
\begin{tikzpicture}[node distance=10pt]
\node[draw, rounded corners, fill=red!30]          (start0)   {Simulate 1D tensionless planar bilayer to get $\omega^{1d}_{A}$, $\omega^{1d}_{B}$ and $z_{c}$};
\node[draw, rounded corners, right=51pt of start0, fill=red!30]          (end0)   {End};
  \node[draw, below=of start0, fill=green!30]          (start)   {Generate 2D initial fields $\omega^{2d}_{A}$, $\omega^{2d}_{B}$ and input $a$, $T$, $n_{cp}$, $\epsilon_{wid}$, $r_A$};
  \node[draw, below=of start, fill=green!30]           (step 1)  {Solve the MDEs for $q^{\pm}_{A},q^{\pm}_{B}$, $q^{h}_{A}$};
  \node[draw, below=of step 1, fill=green!30]          (step 2)  {Calculate the order parameters $\phi_{A}$, $\phi_{B}$};
  \node[draw, right=80pt of step 1, fill=blue!30]          (compare)  {Compare $F_{ex}$ and $F$};
  \node[draw, diamond, aspect=2, below=of step 2, fill=green!30]
  (choice)  {Self-consistent};
  \node[draw, left=20pt of choice, fill=green!30]     (step x)  {Update the $\omega^{2d}_{A}$, $\omega^{2d}_{B}$};
   \node[draw, right=30pt of choice, fill=blue!30]     (step x1)  {Calculate the excess free energy $F_{ex}$};
  \node[draw, below=20pt of choice, fill=green!30]
  (end)  {Output $\omega^{2d}_{A}$, $\omega^{2d}_{B}$, $\phi_{A}$, $\phi_{B}$ };
  \node[draw, right=20pt of end, fill=blue!30]      (goon)  {Extract the bilayer profile $\Gamma$};
    \node[draw, right=25pt of goon, fill=blue!30]      (goon1)  {Calculate the Helfrich energy $F$};
    \draw[->] (start0)  -- (start);
    \draw[->] (compare)  -- (end0);
  \draw[->] (start)  -- (step 1);
  \draw[->] (step 1) -- (step 2);
  \draw[->] (step 2) -- (choice);
  \draw[->] (choice) -- node[left]  {Yes} (end);
  \draw[->] (choice) -- node[above] {No}  (step x);
   \draw[->] (end) -- (step x1);
  \draw[->] (step x) -- (step x|-step 1) -> (start);
  \draw[->] (end) -- (goon);
  \draw[->] (goon) -- (goon1);
  \draw[->] (goon1) -- (compare);
  \draw[->] (step x1) -- (compare);
\end{tikzpicture}
\caption{Procedure of the numerical simulation. Green for the SCF part, and blue for the Helfrich part.}
\label{fig:flow chart}
\end{figure*}

To verify the accuracy of the Helfrich model to predict the free energy of bilayers with general shapes beyond cylindrical and spherical geometries, we extract the elastic constants analytically, follow \cite{Cai2020Elastic}, solve the SCF model equipped with the constraint term $I_{\psi_{i}}$ numerically, and then compare their excess free energies. Although the constraints term is general, this paper only focuses on periodic cylindrical bilayers. With properly assigning several constrain points $\mathbf{r}_i$, the system will adjust the bilayer profile to satisfy the SCF equations. Since the bilayer has two interfaces, while the Helfrich model only accepts one surface, one needs to introduce a procedure to merge the two interfaces as one surface $S$, which will be detailed later. As we are considering periodic cylindrical membranes, the surface $S$ can be simplified as a two-dimensional curve $\Gamma$ parameterized as
\begin{equation}
    \Gamma = \{
        (x(t), y(t)): ~t\in[0,T]
        \},
\end{equation}
where $T$ is the period along $x$-axis. Then the local curvature of $\Gamma$ can be calculated and the Helfrich energy becomes the following:
\begin{align}\label{eq: Helfrich}
    {F}[\Gamma]
    =
    &\int_{0}^{T}
        \frac{8 \kappa_{M} c^2(t) + \kappa_{1} c^4(t) }
        {16\sqrt{x^{\prime 2}(t) +y^{\prime 2}(t)}} \mathrm{d} t,
\end{align}
where $c^2 = {\left(x^{\prime} y^{\prime \prime}-x^{\prime \prime} y^{\prime}\right)^{2}}/{\left(x^{\prime 2}+y^{\prime 2}\right)^{3}}$ is the square of the local curvature. The relationship between $F_{ex}$ and ${F}$ will be examined in the next sections, where the involved numerical procedure is shown in Fig.~\ref{fig:flow chart}.

\section{Method}
\label{sec:Method}

\subsection{Numerical method to solve the SCF model}
\label{subsec:Numerical algorithm of SCF}

We assume that the periodic cylindrical membranes have a period $T$ along the $x$-axis, and the system tends towards the bulk phase on both sides of the $y$-axis. For this setting, the simulation can be approximately reduced to solve a periodic boundary problem in both the $x$ and $y$-axes, where the period on the $y$-axis should be large enough to ensure that the concentrations at the boundary are close to the bulk phase. The computational domain is denoted as $[0, T] \times [-L_y, L_y]$, where $L_y$ is larger and its assignment will be discussed in Subsection~\ref{subsec:Effect of numerical parameters on the energy calculation of the bilayer}. The most time-consuming part of the SCF model is solving the MDEs~(\ref{eq: propagator eqs1}-\ref{eq: propagator eqs2}) with given the fields $\omega_A$ and $\omega_B$. Here, we employ the Fourier spectral method and the fast Fourier transform to accelerate the computation \cite{2008spectral_method, 2013spectral_method}. The calculation steps are summarized as follows: (1) transform Eqs.~(\ref{eq: propagator eqs1}-\ref{eq: propagator eqs2}) into ordinary differential equations using the Fourier transform, (2) solve these equations directly using the given initial conditions, and (3) obtain the solution by applying the inverse Fourier transform. The fields $\omega_A$ and $\omega_B$ are updated by the Picard iteration and accelerated by the Anderson iteration \cite{noolandi1996theory, thompson2004improved} until the fields and concentrations are self-consistent, \emph{i.e.}, the SCF equations Eq.~(\ref{eq: Discrete SCF equations 1}-\ref{eq: Discrete SCF equations 4}) are satisfied.

\subsection{Constraint points and initial fields}
\label{subsec:Choosing constraint points and initial values}

Note that the constraint points are introduced to guide the bilayer to some desired shapes, meanwhile, the initial fields should be assigned accordingly. In our simulation, we chose the standard sinusoidal function as a reference to set the initial fields and constraint points.

\begin{figure}[htb!]
    \center
    \includegraphics[width=9cm]{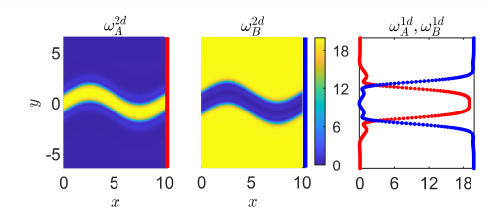}
    \caption{Initial value of the fields with $a=1$ and $T=10$. The one-dimensional fields $\omega_A^{1d}$ (red line) and $\omega_B^{1d}$ (blue line) are used to set $\omega_A^{2d}$ and $\omega_B^{2d}$. }
    \label{fig: Initial_value}
\end{figure}

Let $f(x)$ be the following sinusoidal function,
\begin{equation*}
    f(x) =  a\sin (\frac{2\pi}{T}x),
\end{equation*}
with a period of $T$ and an amplitude of $a$. After solving the SCF model to obtain the fields $\omega_A^{1d}$ and $\omega_B^{1d}$ of a one-dimensional planar bilayer, the initial two-dimensional fields $\omega_A^{2d}$ and $\omega_B^{2d}$ are set as a shift of $\omega_A^{1d}(y-f(x))$ and $\omega_B^{1d}(y-f(x))$. The detailed procedure is provided in Appendix~\ref{sec:Method of designing initial values} and an example of this construction is shown in Fig.~\ref{fig: Initial_value}.

Moreover, the constraint points $\mathbf{r_i}$ are chosen from a parallel curve of $y=f(x)$, as shown in Fig.~\ref{fig:constraint_points_interface_surface}(a). The parallel curve is an extension of $y=f(x)$ along the normal vector $\mathbf{n}=\tfrac{1}{\sqrt{1+f'(x)^2}}(-f'(x),1)$. Therefore, the constraint points are of the form $(x,f(x)) + d \mathbf{n}$, where $d$ is half of the planar bilayer's thickness.

\subsection{Extracting the bilayer profile}
\label{subsec:Extracting the bilayer profile}

After solving the SCF model with the constraint term, we extract the profile of the bilayer from the convergent concentrations $\phi_A$ and $\phi_B$. A natural way is to use the level set of $\phi_A$ or $\phi_B$. However, the bilayer we considered has two interfaces, which means that the level set, such as $\phi_A=\phi_B=0.5$, contains two curves. Since there isn't a standard way to merge the two interfaces into one, we propose several methods to determine the profile and examine the differences later.

\begin{figure*}[!hbpt]
    \center
    \includegraphics[width=0.3\textwidth]{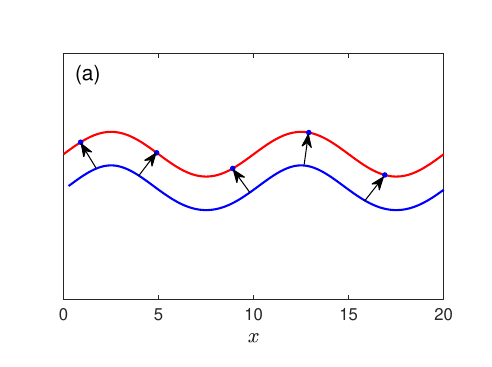}
    \includegraphics[width=0.3\textwidth]{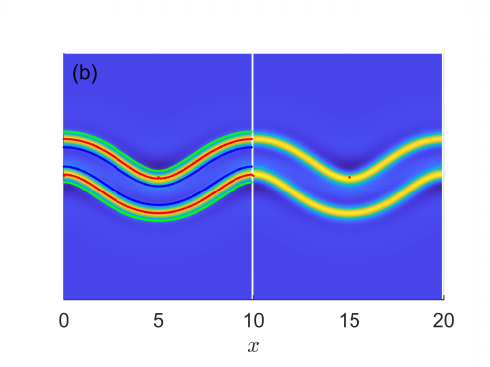}
    \includegraphics[width=0.3\textwidth]{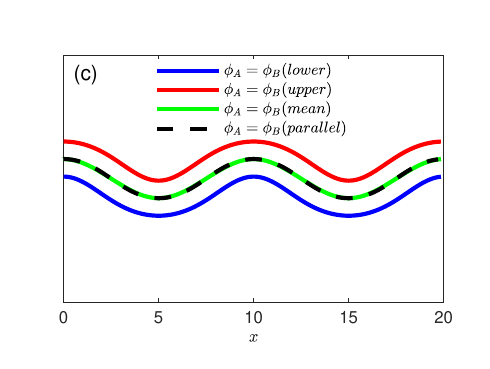}
    \caption{The constraint points and extracted interfaces. (a) The constraint points lie on a parallel surface of $y=f(x)$. (b) Level set of $\phi_A$. The green, red, and blue curves represent $\phi_{A}=0.1, 0.5$, and $0.95$, respectively. The background is colored by the Lagrange multiplier $\xi$. (c) Merge the lower and upper curves into one curve.}
    \label{fig:constraint_points_interface_surface}
\end{figure*}

The methods include two steps: (1) extracting the upper and lower interface curves, $\Gamma_+$ and $\Gamma_-$, parameterized as $r_+(t)$ and $r_-(t)$, respectively, and (2) merging the two curves $\Gamma_\pm$ into one curve, $\Gamma$. It is natural to determine $\Gamma_\pm$ as a level set of $\phi_A$, such as $\phi_A=0.5$. However, the second step is subtle. A simple method can take either $\Gamma_+$ or $\Gamma_-$ as $\Gamma$, and another simple method takes the mean value of $r_+(t)$ and $r_-(t)$ as the parameterization for $\Gamma$. Continuing with the concept of parallel surfaces, we present a third method: taking $\Gamma$ as the curve that is equidistant from $\Gamma_+$ and $\Gamma_-$. The detail is provided in Appendix~\ref{sec:Method of extracting the interface energy} and the procedure is shown in Fig.~\ref{fig:constraint_points_interface_surface}(b,c).

\section{Results and discussion}
\label{sec:Results and discussion}

Now we present the numerical results where the default model parameters are chosen from Cai et al \cite{Cai2019Elastic} as $\chi N=20$, $f_A=f_B=0.5$. The chemical potential $z_c$ is adjusted so that the planar bilayer is tensionless. The default number of discretization points in space is taken as $N_{x}=100$ and $N_{y}=200$, and the number of discretization points in the chain length is $N_{s}=300$. The default numerical parameters are enough to make the results have high accuracies. In addition, the self-consistent fields are updated until the error is less than $10^{-6}$.

\subsection{Effect of numerical parameters on the energy calculation of the bilayer}
\label{subsec:Effect of numerical parameters on the energy calculation of the bilayer}

In this subsection, we will discuss the effect of numerical parameters on the calculated free energy of the bilayer. The numerical parameters include the number of discretization points or grid size, $N_x$, $N_y$, and $N_s$, for the computational domain and the chain length, respectively, the truncated domain size $L_y$ in the $y$-axis, the number of constraint points, $n_{CP}$, and the Gaussian constraint width $\varepsilon$. For the constraint of width, we define a ratio $\epsilon_{wid} = \varepsilon/2d$ by taking the bilayer's thickness $2d$ as a reference width.

\par
\begin{figure}[!h]
    \centering
    \includegraphics[width=8.3cm]{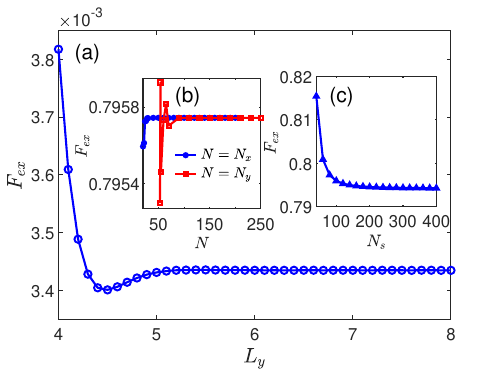}
    \caption{The effect of $L_y$, $N_x$, and $N_y$ on the calculated excess free energy. Parameters: $T=10$, $n_{CP}=2$, $\epsilon_{wid}=1.5$. $a=0$ and 1 for (a) and (b,c), respectively.}
    \label{fig:Dcut_nxny}
\end{figure}

\textbf{Grid size and domain truncation.} It is expected that more discretization points will lead to high accuracy, however, the computation burden is also large. To facilitate the choice of parameters, we compare the free energies calculated under various numerical parameters. Fig.~\ref{fig:Dcut_nxny} shows the free energy of some demo bilayers as a function of $L_y$, $N_x$, $N_y$, and $N_s$, respectively. The energy difference is negligible when $L_y$ is larger, for example, $L_y \ge 5.5$, as the concentrations near the boundary are close enough to the bulk phase. In later calculations, when considering the amplitude $a$ of the bilayer, we will choose $L_{y}=5.5+a$. In addition, the effect of $N_x$, $N_y$, and $N_s$ is also negligible when they are large. This implies that our default setting of $N_x=100$, $N_y=200$, and $N_s=300$ is already sufficiently large.

\par
\begin{figure}[htb!]
    \center
    \scriptsize
    \begin{tabular}{cc}
    \includegraphics[width=7cm]{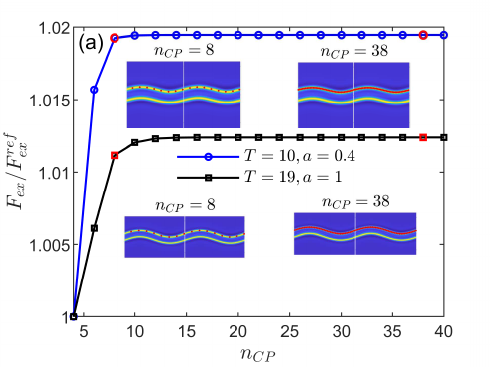}\\
    \includegraphics[width=7cm]{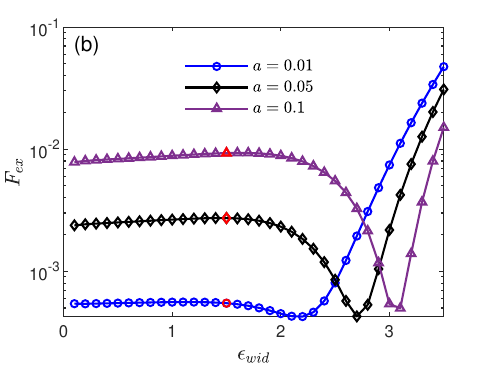}
    \end{tabular}
    \caption{(a) The free energy with a different number of constraint points. $F_{ex}^{ref}$ represents the excess free energy at $n_{CP}=4$. (b) The free energy with different constraint width ratio $\epsilon_{wid}$. The parameters $T=10$ and $n_{CP}=2$.}
    \label{fig:constraint_points_rc}
\end{figure}

\textbf{Constraint points and constraint width.} Note that the constraint term aims to stabilize bilayers with the desired shape. Using more constraint points is expected to take the membranes close to the target shape, but this will limit the membrane's self-assembly flexibility. Fig.~\ref{fig:constraint_points_rc}(a) shows the free energy as a function of the number of constraint points. It implies that an increase in the number of constraint points will result in higher free energy, while a smaller number of constraint points is enough to guide the self-assembly of membrane shapes. Therefore, we only use two constraint points, \emph{i.e.}, $n_{CP}=2$, in subsequent simulations. This allows the membrane to flexibly adjust its interfaces.

Since the membrane itself has a certain thickness, the non-zero constraint width ratio $\epsilon_{wid}$ will affect the final profile of the self-assembled bilayer. Fig.~\ref{fig:constraint_points_rc}(b) shows the effect of $\epsilon_{wid}$ on the free energy of bilayers with different amplitudes. It can be observed that the free energy is slightly changed when the ratio $\epsilon_{wid}$ is small. To improve the stability of the numerical simulation, we will set $\epsilon_{wid}=1.5$ in subsequent simulations.

\subsection{Effect of the constraint amplitude and period}
\label{subsec:Effect of the constraint amplitude}

Now we start to examine the accuracy of the Helfrich model to predict the excess free energy of periodic cylindrical bilayers with different amplitude $a$ and period $T$. As mentioned previously, we consider several methods to determine the profile of the bilayer before evaluating the Helfrich model.

\begin{figure}[htb!]
    \center
    \scriptsize
    \begin{tabular}{cc}
    \includegraphics[width=8.3cm]{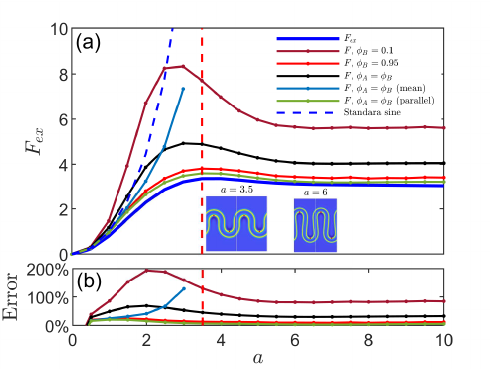}
    \end{tabular}
    \caption{(a) The excess free energy $F_{ex}$ and $F$ vary with the amplitude $a$ at $T=10$. (b) The relative errors in Helfrich's predicted energy vary with the amplitude $a$.}
    \label{fig:a_F_Hefrich}
\end{figure}

Fig.~\ref{fig:a_F_Hefrich} shows a typical example of bilayers with a fixed period $T=10$, where the free energy and its relative error are given as a function of the bilayer amplitude $a$. The elastic constants in the Helfrich model are obtained by the asymptotic expansion method \cite{Cai2020Elastic}. At first, glance, directly taking one level set curve of $\phi_B=0.1$ or $\phi_A=\phi_B=0.5$, or mean the two-level set curves of $\phi_A=\phi_B$ as the shape of the bilayer, will lead to a large margin of error, especially when the amplitude is larger. Particularly, the direct mean of the two-level set curves is not feasible when $a$ is larger than 3, as the curves are not in the form of $y=f_\pm(x)$. However, using the mean parallel curve as the profile of the bilayer could have better accuracy. In addition, taking the level set curve of $\phi_B=0.95$ also gives good accuracy as it is close to the mean parallel curve.

Since we only used two constraint points, the self-assembled bilayer is not necessarily a standard sinusoidal shape. Fig.~\ref{fig:a_F_Hefrich} also gives the Helfrich energy of sinusoidal curves which is larger than that of the self-assembled bilayers whose profiles are shown in Fig.~\ref{fig:a_F_Hefrich} for $a=3.5$ and $a=6$. This agrees with the assumption that the system could adjust the bilayer geometry to decrease the free energy. It is worth noting that the excess free energy of the bilayer remains almost unchanged when $a$ is very large. This is because the membrane is flat for a large part, which has small contributions to the whole energy.

\begin{figure}[htb!]
    \center
    \scriptsize
    \begin{tabular}{cc}
    \includegraphics[width=8.cm]{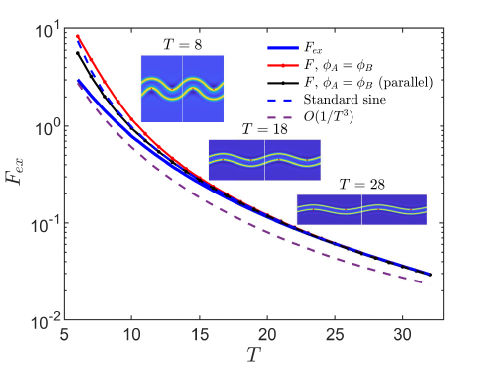}
    \end{tabular}
    \caption{The excess free energy $F_{ex}$ and $F$ vary with the period $T$ with $a=1$.}
    \label{fig:T_F_Hefrich}
\end{figure}

Next, we turn to change the period $T$ and fix the constraint amplitude $a=1$. Fig.~\ref{fig:T_F_Hefrich} shows the free energy as a function of $T$ and some bilayer profiles. Since the curvature tends to zero when the period tends to infinity, the free energy converges to the surface tension which is zero as we are considering the tensionless bilayer. Helfrich's energy of standard sinusoidal curves with corresponding amplitude are also compared, which are close to the SCF calculations when $T$ is large. This indicates that the prediction of the Helfrich model is accurate when the curve of the periodic cylindrical bilayer is small and the excess free energy is of the order $1/T^{3}$.

\subsection{Effect of the interaction strength}
\label{subsec:Effect of the interaction strength}

\begin{figure}[htb!]
    \center
    \scriptsize
    \begin{tabular}{cc}
    \includegraphics[width=7cm] {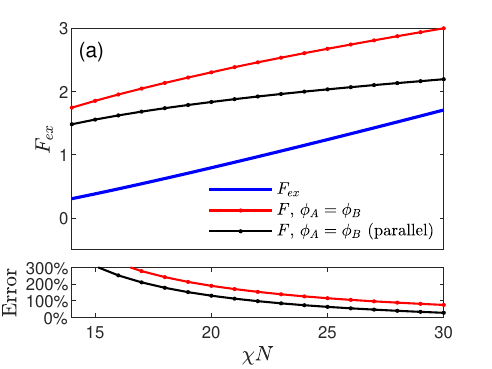}\\
    \includegraphics[width=7cm]{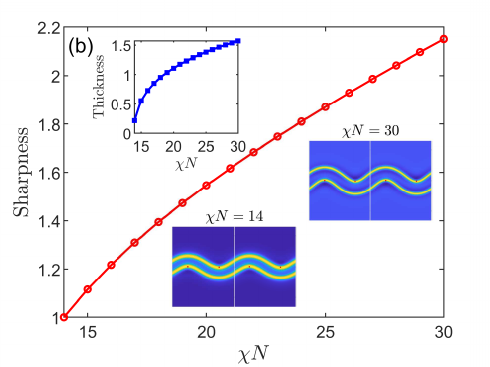}
    \end{tabular}
    \caption{(a) The free energy and (b) the sharpness of bilayers as a function of $\chi N$. The parameters $a=1$ and $T=10$.}
    \label{fig:a_chiNProfile}
\end{figure}

Here, we explore the effect of the interaction strength $\chi N$. As an example, we take $a=1$ and $T=10$, and show the free energy as a function of $\chi N$ in Fig.~\ref{fig:a_chiNProfile}. It is observed that the free energy of bilayers increases almost linearly, which is expected since the bending modulus is almost linearly dependent on $\chi N$ \cite{Li2013Elastic, Cai2020Elastic}. It is interesting to note that the relative prediction error of the Helfrich model decreases when $\chi N$ increases. This might be because the bilayer interfaces become sharp and dominate the system. Fig.~\ref{fig:a_chiNProfile}(a,b) gives the bilayers' sharpness, characterized by $\int |\nabla \phi_A(\mathbf{r})|^2 d \mathbf{r}$ and taken the value at $\chi N=14$ as a reference, which is almost linearly increasing. The bilayer's average thickness is also provided, where the thickness is the distance between two corresponding points on the two interfaces of bilayers. In addition, the shape of several bilayers is also illustrated, which hardly changes for different $\chi N$.

\subsection{Periodic cylindrical bilayers}
\label{subsec:Periodic cylindrical bilayers}

\begin{figure}[htb!]
    \center
    \scriptsize
    \begin{tabular}{cc}
    \includegraphics[width=7.0cm]{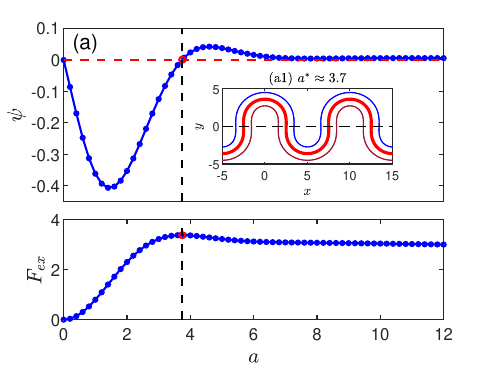}\\
    \includegraphics[width=7.0cm]{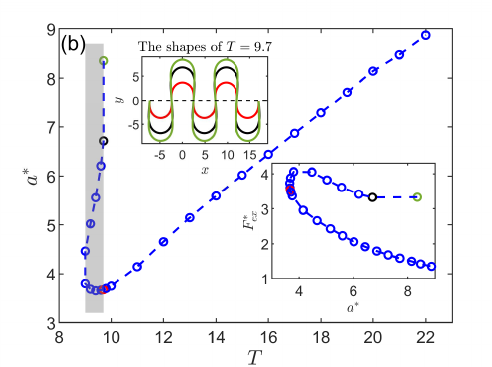}
    \end{tabular}
    \caption{(a) The constraint Lagrange multiplier and free energy of the bilayer with a period $T=10$. (b) The critical amplitude $a^*$ as a function of the period $T$.}
    \label{fig:T_psi}
\end{figure}

During the numerical simulations, we unexpectedly discovered that there are some equilibrium shapes that satisfy the SCF equations without constraints, \emph{i.e.}, all Lagrange multipliers $\psi_i$ in Eq.~(\ref{eq:free energy}) are zero. A typical example of $T=10$ is shown in Fig.~\ref{fig:T_psi} by giving the Lagrange multiplier $\psi_i$ as a function of the constraint amplitude $a$. Note that the Lagrange multiplier $\psi_i$ vanishes near $a=a^{*} \approx 3.7$ and the free energy of the bilayer with this critical amplitude achieves the local maximum. The profile of this special is shown in Fig.~\ref{fig:T_psi}(a1). It is worth noting that this special shape was observed experimentally and subsequently analyzed theoretically as a solution to the shape equations \cite{harbich1984swelling, PhysRevE.53.4206, Tu2014Recent}.

Here, we further illustrate that the special bilayer shape is universal for bilayers with different periods. By examining the constraint Lagrange multipliers, we plot the critical amplitudes $a^*$ as a function of the periods $T$ in Fig.~\ref{fig:T_psi}(b). The free energy of bilayers with these critical amplitudes is also provided. When the period $T$ is large, such as $T\ge 10$, the critical amplitude $a^*$ is almost linearly dependent on the period $T$. In addition, the critical shape is the same in the case of $T=10$ where only the scale of the shape changes for different $T$.

However, when $T$ is small, such as $T<8.9$, there isn't any critical amplitude. The cases of period $T$ between 8.9 and 9.7 seem interesting as there are more than one critical amplitudes $a^*$. For example, there are three critical amplitudes $a^*$ for the case of $T = 9.7$ as shown in Fig.~\ref{fig:T_psi}(b). During these multiple critical shapes, the bilayers with large amplitudes have lower free energies.

\subsection{Asymmetic bilayers}
\label{subsec:Effect of the ratio of sinusoidal symmetry}
Until now, we have been simulating symmetric periodic cylindrical bilayers. However, there are asymmetric bilayers in experiments \cite{1993Theory, 2007Structure}, which are known as the ripple phases of membranes. Here, we asymmetrically assign the constraint points and examine whether the asymmetric bilayers can be in equilibrium states. The two constraint points are located at $x= \pm r_A T$ periodically, where $r_A$ is the ratio of asymmetry and $r_A=0.5$ will lead to symmetric constraint points.

\begin{figure}[htb!]
    \centering
    \scriptsize
    \begin{tabular}{cc}
    \includegraphics[width=8.3cm]{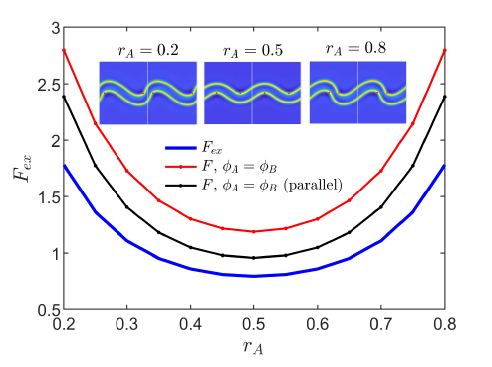}
    \end{tabular}
    \caption{Free energy of asymmetric bilayers with amplitude $a=1$ and period $T=10$.}
    \label{fig:rA_F_Hefrich}
\end{figure}

Some examples of asymmetric bilayers are shown in Fig.~\ref{fig:rA_F_Hefrich} where the free energy and profiles are given. When the value of $r_A$ is either small or large, the profile of the bilayer differs significantly from that of the symmetric bilayer with $r_A=0.5$. For the case of $r_A \neq 0.5$, we can also obtain equilibrium bilayers by repeating the procedure described in the previous subsection to determine the critical amplitude $a^*$. Unfortunately, the equilibrium bilayers obtained are symmetric and have the same shape as the bilayer shown in Fig.~\ref{fig:T_psi}(a1). Note that ripple phases in experiments are observed for lipid bilayers, where the molecules involve rigid or semi-flexible chains. Our simulations suggest that the asymmetric periodic cylindrical bilayers are not equilibrium states for flexible bilayers.

\section{Conclusion}
\label{sec:Conclusion}

In this paper, we simulate periodic cylindrical bilayers within the SCF framework and verify the accuracy of the Helfrich model to predict the free energy of self-assembled bilayers. The bilayers are stabilized by introducing some constraint points and the effect of the constraints on the self-assembled bilayers is systematically examined. Numerical results show that the method of reducing the bilayer as a surface is crucial to using the Helfrich model, especially when the bilayer's curvature is large.

Furthermore, we have identified certain critical bilayers that represent equilibrium states of the SCF model without any constraints. However, the equilibrium bilayers obtained are symmetric, despite the asymmetric assignment of constraint points. This suggests that we should extend the simulation to include periodic liquid-crystalline bilayers, where asymmetric ripple phases have been observed in experiments. A systematic study in this area is intriguing, and we will consider it for future research.

\begin{acknowledgments}
YQ Cai is supported by the National Natural Science Foundation of China (grant no. 12201053). YN Di is partially supported by the National Natural Science Foundation of China (grant no. 12271048), and Guangdong Key Laboratory.
\end{acknowledgments}

\appendix

\section{The bulk phase}
\label{sec:Bulk phase}

The SCF equations have multiple solutions, including the constant solution, which is also known as the bulk phase. In our numerical simulation, the bulk phase is used as the reference state, where its free energy density is given by:
\begin{align}
    \frac{N \mathcal{F}_{b u l k}}{k_{B} T \rho_{0} V}
    =
    \ln \left(1-\phi_{\text {bulk }}\right)
    +\chi N f_{B}^{2} \phi_{\text {bulk }}^{2}-1,
\end{align}
where $\phi_{\text {bulk }}$ is the bulk copolymer concentration determined by the following equation:
\begin{align}\label{eq:phi_bulk}
    \mu_{c}
    =
    \ln \Big( \frac{\phi_{\text {bulk }}}{ 1-\phi_{\text {bulk }}} \Big)
    +\chi N f_{B}\left(1-2 f_{B} \phi_{\text {bulk }}\right).
\end{align}
When Eq.~(\ref{eq:phi_bulk}) has more than one solution, the one with the lowest free energy density is chosen as the bulk phase.

\section{Design the initial fields}
\label{sec:Method of designing initial values}

In the SCF model, the initial guess of the fields is crucial for efficient simulations. To obtain the periodic cylindrical bilayers, we construct the initial fields based on the results of the one-dimensional planar bilayers. The specific steps are as follows:
\begin{enumerate}
    \item [(1)] Calculate the one-dimensional fields $\omega_A^{1d}(x)$ and $\omega_B^{1d}(x)$ for tensionless bilayers. Here the chemical potential $z_c$ is adjusted such the bilayer's excess free energy is zero.

    \item [(2)] Construct the two-dimensional simulation grid. The computational domain is $\left[0, T\right]\times \left[-L_y, L_y\right]$, which is uniformly discretized as is $N_x \times N_y$ grid points.

    \item [(3)] Construct initial guess of the two-dimensional fields $\omega_A^{2d}(x,y)$ and $\omega_B^{2d}(x,y)$. Denote the two points that satisfying $\omega_A^{1d}(x)= \omega_B^{1d}(x)$ as $x_1$ and $x_2$, and define their mean value as $\bar x=(x_1+x_2)/2$. Then the two-dimensional fields are assigned as
    \begin{align*}
        \omega_A^{2d}(x,y)= \omega_A^{1d}(y-f(x)+\bar x), \\
        \omega_B^{2d}(x,y)= \omega_B^{1d}(y-f(x)+\bar x).
    \end{align*}
    The evaluation of $\omega_A^{1d}$ and $\omega_B^{1d}$ is performed using linear interpolation with the values at one-dimensional grid points.

\end{enumerate}

\section{Extract the interfaces}
\label{sec:Method of extracting the interface energy}

We determine the shape of the bilayer based on its order parameters. The following methods are considered:
\begin{enumerate}
    \item [(1)] Take one unilateral level set of $\phi_{A}$, such as the upper part of the set $\phi_{A}=0.5$, as the bilayer's shape.

    \item [(2)] Directly average the two interfaces such that $\phi_{A}=\phi_{B}$ along $y$-axis as the bilayer's shape.

    \item [(3)] Take the parallel median of the two interfaces as the bilayer's shape.
\end{enumerate}
Here, the parallel median of two interface curves $\Gamma_\pm$ parameterized as $r_\pm(t)$ is the curve $\tilde r (t)=(r_+(t)+r_-(t'))/2$, where $t$ and $t'$ are paired such that the tangent lines of $\Gamma_\pm$ at $t$ and $t'$, respectively, are parallel.

\bibliographystyle{acm}
\nocite{*}
\bibliography{refs}

\end{document}